%% file: 0main.tex
\documentclass[sigconf]{acmart}

\AtBeginDocument{%
  \providecommand\BibTeX{{%
    \normalfont B\kern-0.5em{\scshape i\kern-0.25em b}\kern-0.8em\TeX}}}

\setcopyright{acmcopyright}

\copyrightyear{2023}
\acmYear{2023}
\setcopyright{acmlicensed}\acmConference[WWW '23]{Proceedings of the ACM Web Conference 2023}{May 1--5, 2023}{Austin, TX, USA}
\acmBooktitle{Proceedings of the ACM Web Conference 2023 (WWW '23), May 1--5, 2023, Austin, TX, USA}
\acmPrice{15.00}
\acmDOI{10.1145/3543507.3583418}
\acmISBN{978-1-4503-9416-1/23/04}

\usepackage{booktabs} 
\usepackage{amsmath}
\usepackage{amsfonts}
\usepackage{comment}
\usepackage{footmisc}
\usepackage{mathrsfs}
\usepackage{graphicx}
\usepackage{subfigure}
\usepackage{multirow}
\usepackage{algorithm}
\usepackage{algorithmic}
\usepackage{multicol}  
\usepackage{enumitem}
\usepackage{array}
\usepackage{float}
\usepackage{hyperref}
\hypersetup{hidelinks}
\usepackage{bm}
\usepackage{dsfont}

\usepackage{color}

\begin{document}
\title{User Retention-oriented Recommendation \\ with Decision Transformer}

\author{Kesen Zhao}
\affiliation{\institution{City University of Hong Kong}}
\email{kesenzhao2-c@my.cityu.edu.hk}

\author{Lixin Zou$^\dagger$}
\affiliation{\institution{Wuhan University}}
\email{zoulixin15@gmail.com}

\author{Xiangyu Zhao$^\dagger$}
\affiliation{\institution{City University of Hong Kong}}
\email{xianzhao@cityu.edu.hk}

\author{Maolin Wang}
\affiliation{\institution{City University of Hong Kong}}
\email{MorinWang@foxmail.com}

\author{Dawei Yin}
\affiliation{\institution{Baidu Inc.}}
\email{yindawei@acm.org}

\thanks{
$^\dagger$ Corresponding author. 
}
\begin{abstract}
Improving user retention with reinforcement learning~(RL) has attracted increasing attention due to its significant importance in boosting user engagement. 
However, training the RL policy from scratch without hurting users' experience is unavoidable due to the requirement of trial-and-error searches. 
Furthermore, the offline methods, which aim to optimize the policy without online interactions, suffer from the notorious stability problem in value estimation or unbounded variance in counterfactual policy evaluation. 
To this end, we propose optimizing user retention with Decision Transformer~(DT), which avoids the offline difficulty by translating the RL as an autoregressive problem. 
However, deploying the DT in recommendation is a non-trivial problem because of the following challenges: 
\textbf{(1)} deficiency in modeling the numerical reward value; 
\textbf{(2)} data discrepancy between the policy learning and recommendation generation; 
\textbf{(3)} unreliable offline performance evaluation. 
In this work, we, therefore, contribute a series of strategies for tackling the exposed issues. 
We first articulate an efficient reward prompt by weighted aggregation of meta embeddings for informative reward embedding. 
Then, we endow a weighted contrastive learning method to solve the discrepancy between training and inference. Furthermore, we design two robust offline metrics to measure user retention. 
Finally, the significant improvement in the benchmark datasets demonstrates the superiority of the proposed method. 
The implementation code is available at \url{https://github.com/kesenzhao/DT4Rec.git}.
\end{abstract}






\makeatletter
\def\@ACM@checkaffil{
    \if@ACM@instpresent\else
    \ClassWarningNoLine{\@classname}{No institution present for an affiliation}%
    \fi
    \if@ACM@citypresent\else
    \ClassWarningNoLine{\@classname}{No city present for an affiliation}%
    \fi
    \if@ACM@countrypresent\else
        \ClassWarningNoLine{\@classname}{No country present for an affiliation}%
    \fi
}
\makeatother

\maketitle
\input{1Introduction}
\input{2Framework}
\input{3Experiments}
\input{4RelatedWork}
\input{5Conclusion}
\bibliographystyle{ACM-Reference-Format.bst}
\bibliography{9Reference}
\end{document}

%% file: 1Introduction.tex
\section{Introduction}

Sequential recommender systems~(SRSs), which model users' historical interactions and recommend potentially interesting items for users, have received considerable attention in both academia and industry due to their irreplaceable role in the real world systems, e.g., movie recommendation in Netflix\footnote{\url{https://www.netflix.com/}}, and E-commerce recommendation in Amazon\footnote{\url{https://www.amazon.com/}}. 
The success of SRSs heavily relays on users' engagement on the platform, which, to some extent, is reflected by users' immediate feedback, liking clicks~\cite{das2007google,zhao2019deep}. However, these immediate feedback can not completely reveal users' preferences~\cite{yi2014beyond}. 
For example, some items with eye-catching titles and covers but low-quality content may attract users' clicks and further break users' trust in the platform~\cite{wu2017returning}. 
Therefore,  it is essential to optimize users' long-term engagement at the platform~\cite{zou2019reinforcement}, liking user retention, which is a preferable indicator of user satisfaction.



As the tool for optimizing the long-term/delayed metrics~\cite{lin2021survey}, reinforcement learning~(RL) has been widely studied for optimizing user retention in recent years~\cite{chen2021survey}. 
Though they are capable of exploring and modeling users' dynamic interests~\cite{zhang2020deep}, existing RL-based SRSs leave much to be desired due to the offline learning challenge. 
Unlike gaming scenarios, where RL agents achieve great success by trial and error search~\cite{silver2016mastering}, training from scratch in online SRSs is unaffordable due to the risk of losing users by recommending inappropriate items. Therefore, recent attention of the whole community has been paid to offline RL-based SRSs. However, putting offline RL into practice is frustrating in both value-based and policy-based methods. For value-based approaches, the notorious instability problem~(i.e., the 'Deadly Triad') pushes the development of model-based mothods~\cite{chen2019generative}. However, due to the vast state space in the recommendation scenario, estimating the transition probability is a problem and further leads to unsatisfactory performance~\cite{ zhao2020whole}. For policy-based methods, the unbounded variance of counterfactual policy evaluation drives the community to clip or discard the counterfactual weights~\cite{chen2019top}, which might lead to inaccurate weight and discourage performance~\cite{xin2022rethinking}.

To explore the potential of RL-based recommendation, we propose to optimize the user retention recommendation with Decision Transformer~\cite{chen2021decision}~(DT), which casts the offline RL as an autoregressive problem and therefore solves the mentioned offline learning challenges. Specifically, DT is required to generate the recommendation under a specific reward, i.e., the user retention and state. When conditional on optimal reward, DT can generate future actions that achieve the desired return in the recommendation stages.
Though DT is promising in the recommendation, applying the DT in SRSs is a non-trivial problem. 
It has the following challenges: \textbf{(1) deficiency in reward modeling.} Reward, as the most crucial part of DT, directly affects the quality of the recommendation. 
However, in DT, translating the reward into embedding ignores its partial order, leading to the deficiency in model training. 
\textbf{(2) discrepancy in recommendation generation.} Under the DT, the recommendation is generated with the maximum reward, which is inconsistent with the diversified reward encountered in the training stages. 
As a result, the model is not capable of utilizing the knowledge of data with smaller rewards~(depicted in Figure~\ref{fig:data}).
\textbf{(3) unreliable performance evaluation.} Though DT solves the problem of offline learning, we still need the importance weighted offline evaluation to measure the performance of the learned policy, which leads to unbounded variance and results in an unreliable evaluation.

To handle these problems, we propose a novel framework DT4Rec, which firstly deploy \textbf{D}ecision \textbf{T}ransformer for \textbf{rec}ommendation. Specifically, 
we generate the reward embedding by auto-weighted aggregating the meta-embeddings generated from discretized reward value, which maintains partial order relationship between rewards.
Then, we introduce weighted contrastive learning to remedy the discrepancy between inference and training, which leverages the smaller reward samples by contrasting the large and small reward samples. 
Furthermore, we propose two novel reliable metrics, i.e., the model-based and the similarity-based user retention score, to evaluate the policies all around. 
Compared with the off-policy evaluation methods, it achieves lower variance, i.e., more stability, 
on the performance evaluation~(depicted in Figure~\ref{fig:variance}). 
Our main contributions can be summarized as follows:
\begin{itemize}[leftmargin=*]
    \item We propose a novel Decision Transformer-based SRS model, DT4Rec, which eliminates the difficulty of offline learning with an RL-based recommender system.
    \item We contribute the auto-discretized reward prompt and contrastive supervised policy learning, which effectively cope with the deficiency in reward modeling and discrepancy between training and inference respectively. 
	\item We design the model-based and the similarity-based user retention score. Compared with the off-policy evaluation methods, they can fairly evaluate model performance. 
	\item Experiments on two benchmark datasets illustrate the superiority of our proposed model.
\end{itemize}

%% file: 2Framework.tex
\section{Methodology}
\subsection{Problem Formulation}
In a typical sequential recommendation, we are given a sequence of $M$ user-interested items denoted as  $U_{t}=\{v^t_i\}_{i=1}^M$ in the $t$-th recommendation and seek to predict a set of $N$ items $a_{t}=[v_{t,1},\ldots, v_{t,N}]$ that the user might be interested in. Formally, we aim at generating the recommendation $a^\ast_{t}$, maximizing a specific metric $\Theta$ as
\begin{equation}
a^\ast_t = \mathop{\arg\max}_{a_t} \\ \Theta(a_t|U_t).
\end{equation}
According to the target of sequential recommender, we could instantiate $\Theta$ with the metric of optimizing user immediate feedback~(i.e., prediction accuracy) or long-term user engagement~(e.g., user retention). For user retention, we define it as the number of users logging in the platform over the next $K$ time intervals~(e.g., next $K$ days, next $K$ months)
\begin{equation} 
e_{t}=\sum_{k=1}^{K} \mathds{1}[\text{log-in}_{t+k}],
\end{equation}
where $\mathds{1}$ is the indicator function, $\text{log-in}_{t+k}$ is a binary function that indicates whether the user logs in the platform or not at the next $(t+k)$-th timestep.

    
    
\paragraph{MDP Formulation of Sequential Recommendation}
To optimize the metric $\Theta$ with reinforcement learning, we formally introduce the sequential recommendation under the framework of the Markov Decision Process~(MDP), including a sequence of states, actions, and rewards. Particularly, the state $s_t = U_{t}$ is defined as the historical interactions. The action $a_t$ is the recommendation list, which is generated by a recommendation policy $\pi:s_t \rightarrow a_{t}$. Before every recommendation, $s_t$ will be updated by appending the user's interested items at its end as $s_t = s_{t-1}\oplus\{U_{t}-U_{t-1}\}$. Given a specific evaluation metric $\Theta$, the reward is usually defined as the function of the metric. Particularly, for user retention, the reward is set the same as the $e_t$. With the formally defined $s_t$, $a_t$, $r_t$, user systems' interactions form a trajectory
\begin{equation}
\tau=[s_{1}, a_{1},r_{1},\ldots,s_{t}, a_{t},r_{t}].
\end{equation}
\begin{figure*}[t]
	\centering
	\includegraphics[width=\linewidth]{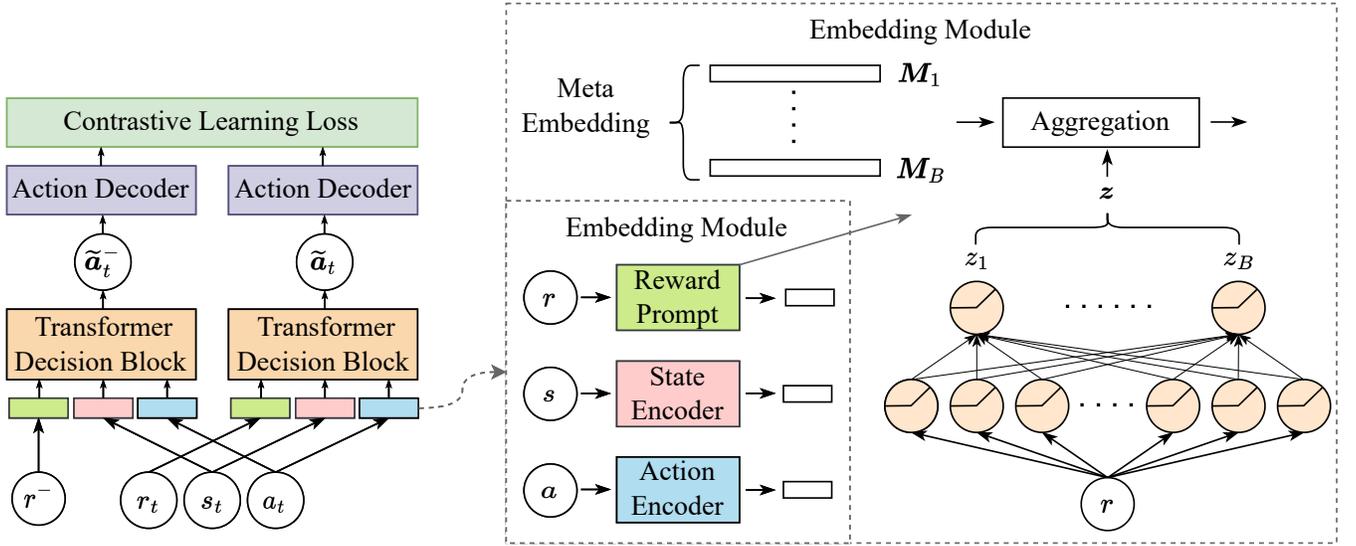}
	\caption{Framework overview of DT4Rec. Negative samples have the same states and actions as positive samples, while rewards are replaced with different values. Positive and negative samples share the same model parameters.}
	\label{fig:Fig1_Overview}
\end{figure*}

\subsection{Decision Transformer based Recommendation}
Unlike traditional frameworks that approximate the state-action value for planning~\cite{ie2019slateq} or directly optimize gradient-based policy function~\cite{chen2021user}, the Decision Transformer translates the RL problem as an autoregression problem by directly predicting the desired action with previous rewards, states, and actions~(refer to Section~\ref{sec_supervised_learning}).  
As an supervised learning task, DT avoids the problems of `Deadly Traid' and unbounded variance and shows superiority in offline learning~\cite{chen2021decision}.
To instantiate the DT in SRSs, we formally disassemble the DT into the following four blocks~(shown in Figure~\ref{fig:Fig1_Overview}): 

\begin{itemize}[leftmargin=*]

\item \textbf{Embedding Module}:
The embedding module aims to map the rewards, states, and actions into dense vectors. Particularly, before translating raw $s_t$, $a_t$ and $r_t$ into embedding, we need to cast the data in the trajectory ordered by reward, state and action to simplify the prediction of action as
\begin{equation} 
\tau'_t=[\widehat{r}_{1},s_{1}, a_{1},\ldots,\widehat{r}_{t}, s_{t},a_{t}],
\end{equation}
where $\widehat{r}_{t}=\sum_{t^{\prime}=t}^{T} e_{t^{\prime}}$ is the cumulative reward, i.e., the return-to-go~\cite{chen2021decision}. $T$ is the round of total recommendation. Then, embedding module sequentially converts the raw feature sequence $\tau'_t$ into feature vector sequence as $\bm{\tau}'_t =[\widehat{\bm{r}}_{1},\bm{s}_{1}, \bm{a}_{1},\ldots,\widehat{\bm{r}}_{t}, \bm{s}_{t},\bm{a}_{t}]$ with the encoder model~(Section~\ref{sec_action_encoder}). Particularly, due to the importance of reward, we distinguish the embedding module for reward as the \textbf{reward prompt}, which is designed to prompt the DT generating desired recommendation lists. 

\item \textbf{Decision Block}:
The decision block is the centre of the model, which transforms the dense vectors of contextual information $\bm{\tau}'_t -\{\bm{a}_t\}=[\widehat{\bm{r}}_{1},\bm{s}_{1}, \bm{a}_{1},\ldots,\widehat{\bm{r}}_{t}, \bm{s}_{t}]$ into context feature $\bm{A}_t$ for generating the desired response in the next time step, where $\widehat{\bm{r}}_{t}$ is the generated reward prompt.  


\item \textbf{Action Decoder}:
Given contextual information $\bm{A}_t$, action decoder is expected to generate a sequence of action $\hat{a}_t$ that matches the ground truth action $a_t$. 
\item \textbf{Supervised Policy Learning}: The goal of supervised policy learning is to minimize the difference between the generated action $\hat{a}_t$ and ground truth $a_t$ with a specific loss function. Therefore, it translates the ordinary RL task into a supervised learning problem. By specifying the optimal reward as the prompt in inference, the DT is desired to recommend the items that can maximize user retention.  

\end{itemize}

\subsection{Auto-Discretized Reward Prompt} 

Reward, as the target of RL, differentiates the performance of different policies through its numerical value. Therefore, the prompts generated by DT should maintain the partial order relationship between rewards, i.e., if two rewards are similar, then the Euclidean distance between their generated prompts is smaller. 

To this end, we propose to generate more efficient prompts by the auto-discretizing method, as shown in Figure \ref{fig:Fig1_Overview}, which consists of discretizing the numerical value and auto-weighted aggregating the meta-embedding learned by MLP. It shares the embedding between similar rewards. Specifically, according to the reward value, we convert it into a weighted score for a batch of $B$ learnable embeddings $\boldsymbol{M}_{b}\in\mathbb{R}$, $\forall b\in[1,B]$ as
\begin{equation} 
\boldsymbol{z}=	\phi \left(\boldsymbol{W} \sigma\left(\boldsymbol{w} \widehat{r}\right) + \alpha \sigma\left(\boldsymbol{w} \widehat{r}\right) \right), 
\end{equation}
where $\boldsymbol{w} \in \mathbb{R}^{1 \times B}$ and $\boldsymbol{W} \in \mathbb{R}^{B \times B}$ are learnable variables. $\sigma$ is the Leaky Relu~\cite{glorot2011deep}. $\phi$ is the Softmax function. Given the weighted score $\boldsymbol{z}$, the reward embedding $\widehat{\boldsymbol{r}}$ is set as the aggregation of meta embedding, which can be formulated as
\begin{eqnarray} 
\widehat{\boldsymbol{r}}=\sum_{b=1}^{B}{\boldsymbol{z}_{b}\boldsymbol{M}_{b}}, 
\end{eqnarray}
where $\boldsymbol{z}_{b}$ is the $b^\text{th}$ element of $\boldsymbol{z}$.
Since the value of $r$ is directly used as the input of the neural network to ensure the partial order between rewards, the similar $\widehat{r}$  will share the similar embedding as long as the neural network is smooth.

\subsection{State-Action Encoder}\label{sec_action_encoder}
The action encoder maps actions to vectors. The challenge lies in the dynamic length of recommender action since the user may interact with different numbers of items each time. 
Therefore, we model the sequence of interaction sequence with GRU, which has shown better performance than LSTM~\cite{hidasi2015session} in modeling the dynamic item sequence for recommender systems. Furthermore, compared with the heavy Transformer~\cite{karita2019comparative} based methods, GRU is a more affordable alternative for balancing the efficiency and effectiveness. 
Specifically, we set the maximum length of sequences to $N$. For those less than $N$, we pad the sequence to $N$ with $\mathbf{0}$ vector~\cite{tan2016improved}. 
Formally, the sequence information can be extracted as
\begin{eqnarray} 
\boldsymbol{H}_{n}&=&GRU_{e}\left(\boldsymbol{v}_{n}, \boldsymbol{H}_{n-1}\right)\\
\boldsymbol{a}_{t}&=&\boldsymbol{H}_{N},
\end{eqnarray}
where $GRU_{e}$ is the recurrent layer, $N$ is maximum sequence length, $\boldsymbol{H}_{n}$ is the hidden state. Additionally, we set the embedding of $a_{t}$ as the hidden state of last timestep, i.e., $\boldsymbol{a}_{t}$.



\subsection{Transformer Decision Block}
Prior efforts have demonstrated the superiority of Transformer and RL in SRS tasks.
On the one hand, Transformer equally treats the element of sequence through self-attention, which avoids information loss of long sequence in RNN and further diminishes the gradient vanishing or explosion in modeling long sequence~\cite{karita2019comparative,zou2021pre,zou2022pre}. 
On the other hand, RL is expert at dynamically modeling users' interest profiles and capturing users' interest shifts. 
To get the best of both worlds, we propose to learn user interests from their interaction trajectories via Transformer Decision blocks~\cite{chen2021decision}.




Specifically, for user retention recommendations, we are required to model the dynamic contextual information for generating the recommender decision, which is similar to the generative task. Therefore, we select the unidirectional Transformer layer as the backbone model for modeling the complex feature interactions, which has shown substantial advantage over existing methods in generative task~\cite{floridi2020gpt}. Furthermore, we employ the skip-connections for mitigating the over-fitting \cite{he2016deep} and feed-forward neural layers for linear-mapping of features \cite{kenton2019bert}. Consequently, the contextual information for recommender decision can be formulated as
\begin{eqnarray} 
\widetilde{\boldsymbol{A}} = FFN\left(MultiHeadAttention\left(\boldsymbol{\tau}' - \{\boldsymbol{a}_t\}\right)\right), 
\end{eqnarray}
where $\widetilde{\boldsymbol{A}}$ is the matrix of predicted action embedding with the $t$-th row as $\widetilde{\boldsymbol{a}}_{t}$ for $t\in[1,T]$. $FFN(x)=GELU(x\boldsymbol{W}_{1}+\boldsymbol{b}_{1})\boldsymbol{W}_{2}+\boldsymbol{b}_{2}$ is the feed-forward neural layer with skip-connection. Here, $GELU$~\cite{hendrycks2016gaussian} is commonly used activation function in Transformer models.
The $MultiHeadAttention$ presents the multi-head self-attentive mechanism defined in~\cite{vaswani2017attention}, which has been proved effective for jointly learning information from different representation subspaces and , therefore, massively deployed in the recommender systems~\cite{kang2018self}.

\subsection{Action Decoder}
Given the contextual information of items interacted with at each time step, action decoder aims to decode sequences of items of interest to users with the GRU~\cite{cho2014learning}. 
The item of current interaction is unknown for decoding. We only use previous interactions and contextual information to predict it as
\begin{eqnarray} 
\widehat{\boldsymbol{v}}_{n}&=&\boldsymbol{v}_{n}  \oplus \widetilde{\boldsymbol{a}}_{t}\\
\widetilde{\boldsymbol{v}}_{n+1}&=&GRU_{d}\left(\widetilde{\boldsymbol{v}}_{n}, \widehat{\boldsymbol{v}}_{n}\right),
\end{eqnarray}
where $\boldsymbol{v}_{n}$ is the embedding of $v_{n}$, $\widetilde{\boldsymbol{v}}_{n+1}$ is the prediction for the item at $n+1$ place, $\oplus$ represents concatenate operations. Since there is no information for predicting the first item, we use `bos' token as the start marker and initialize the $\widetilde{\boldsymbol{v}}_{0}$ randomly. 
The full sequence can be decoded as
\begin{equation}  
\begin{aligned}
\widetilde{\boldsymbol{V}}
&=\text{decoder}\left(\text{bos},\widehat{\boldsymbol{v}}_{1},\ldots, \widehat{\boldsymbol{v}}_{n},\ldots, \widehat{\boldsymbol{v}}_{N-1}\right) \\
&=[\widetilde{\boldsymbol{v}}_{1},\ldots, \widetilde{\boldsymbol{v}}_{n},\ldots, \widetilde{\boldsymbol{v}}_{N}],
\end{aligned}
\end{equation}
where $\widetilde{\boldsymbol{V}}$ is predicted matrix for interacted items sequence at the $t$-th time step, $N$ is maximum sequence length.

Notice that the length of sequences may not be fixed. When predicting,  we do not know the length of sequences in decoder, so we add an `eos' token at the end of each sequence as an end marker and
then still pad it with zeros to a length of $N$. 
The decoder does not continue predicting when it has predicted `eos'.



\subsection{Contrastive Supervised Policy Learning}\label{sec_supervised_learning}
In Decision Transformer, only the maximal reward will be used for inference since it is designed to generate the actions with maximal reward. Therefore, the samples with small rewards might not be fully utilized~(validated in Section \ref{sec_ood_analysis}). In order to fully exploit the knowledge from the samples, we propose to use a weighted contrastive learning approach, which treats actions with small rewards as negative samples to avoid recommending small rewarded actions. Therefore, our objective function consists of two parts, CE loss and weighted contrastive learning loss.

\subsubsection{Weighted Contrastive Loss} 
For each sample, we use same state and action, with different rewards, as negative samples, denoting its predicted action embedding as $\widetilde{\boldsymbol{V}}^{-}$. 
The weighted contrastive learning loss can be formulated as
\begin{equation} 
\mathcal{L}_{\text{CL}} = -\sum_{\widetilde{\boldsymbol{V}}^{-} \in \boldsymbol{\Upsilon}}{\kappa\left(\widetilde{\boldsymbol{V}}^{-}\right) f_{s}\left(\widetilde{\boldsymbol{V}},\widetilde{\boldsymbol{V}}^{-}\right)}, 
\end{equation}
where $\Upsilon$ is the set of negative samples. $f_{s}(\cdot)$ calculates the similarity of two sequences by averaging of embeddings' dot product at every row of the matrices $\widetilde{\boldsymbol{V}}$ and $\widetilde{\boldsymbol{V}}^{-}$. 
$\kappa\left(\widetilde{\boldsymbol{V}}^{-}\right)$ is the weighting hyperparameter set according to the reward value of negative sample, i.e., the weight is inversely proportional to reward. The reason is that smaller rewards make user retention lower and we want to be dissimilar to those ones.

Besides the weighted contrastive loss, we also optimize the DT with the original cross-entropy loss as 
\begin{eqnarray} 
\widehat{\boldsymbol{Y}}&=&\psi\left( \boldsymbol{W}_{v} \widetilde{\boldsymbol{V}} + \boldsymbol{b}_{v} \right) \\
\mathcal{L}_{CE}&=&\textit{CE}\left(\widehat{\boldsymbol{Y}}, \boldsymbol{Y}\right),
\end{eqnarray}
where $\boldsymbol{Y}$ is the label matrix with the one-hot label in every row,
$\widehat{\boldsymbol{Y}}$ is the corresponding predicted label distribution. 
Here, $\textit{CE}$ is the cross-entropy loss function, $\psi$ is the softmax function.  $\boldsymbol{W}_{v}$ and $\boldsymbol{b}_{v}$ are learnable parameters.

Finally, the overall loss function is formulated by combining the cross-entropy loss and weighted contrastive loss as 
\begin{equation} 
\mathcal{L} = \mathcal{L}_{CE} + \beta \mathcal{L}_{CL}, 
\end{equation}
where $\beta$ is a hyper-parameters.  

%% file: 3Experiments.tex
\section{Experiments}
In this section, we conduct experiments on two benchmark datasets, IQiYi and ML-1m, to answer the following research questions.
\begin{itemize}[leftmargin=*]
    \item \textbf{RQ1}: How does the performance of our proposed DT4Rec compare with other state-of-the-art baselines?
    \item \textbf{RQ2}: How does our auto-discretized reward prompt contribute to reward modeling? 
    \item \textbf{RQ3}: How does our contrastive supervised policy learning method mitigate the out-of-distribution (OOD) issue between the recommendation training and inference? 
    \item \textbf{RQ4}: How does stability of our designed evaluation method?
    \item \textbf{RQ5}: How does the generalization capability of our DT4Rec?
\end{itemize}

\begin{table}[t] \center 
\caption{Statistics of datasets. UR is the short for average user retention.}
	\begin{tabular}{@{}c c c c c c@{}}
		\toprule[1pt]
		Datasets & Users & Items & Interactions & UR & Density\\ \midrule
		IQiYi & 3,000,000 & 4,000,000 & 71,046,026 & 4.80 & 5.8e-6 \\ 
		ML-1m & 6,014 & 3,417 & 1,000,000 & 4.13 & 4.84\% \\
		 \bottomrule[1pt]
	\end{tabular}
	\label{tab:datasets}
	\vspace{-3mm}
\end{table}
\subsection{Datasets}
We conduct experiments on two real datasets, iqiyi user retention data (IQiYi) and ML-1m. In Table~\ref{tab:datasets}, we provide details of two datasets. Explicitly, the \textbf{IQiYi dataset}\footnote{\label{note1}\url{http://challenge.ai.iqiyi.com/}} records about 70 million users interactions for 4 million videos, e.g., views and comments, which is a very sparse and noise dataset. Therefore, without losing generality, we select the users with at least 20 interaction records for experiment.
Furthermore, we do not distinguish between the types of interactions and treat them as the identical interactions. 
User retention is set as the average number of days users log in the next week, i.e., $K=7$~(the setting used in WSDM Cup $2022^{\ref{note1}}$). 
The \textbf{ML-1m dataset \footnote{https://grouplens.org/datasets/movielens/1m/}} is a benchmark dataset commonly used in SRSs, which records 6,014 users' scores for 3,417 movies. Since the ML-1m records span a large time scope, we calculate the user retention in the perspective of month.

\subsection{Evaluation}
We use the following evaluation metrics to evaluate the performance of the proposed methods on both immediate feedback~(i.e., prediction accuracy) and long-term engagement~(i.e., user retention). Specifically, these metrics are defined as follows.

\subsubsection{Prediction Accuracy.} 
To assess prediction accuracy, we use four widely used common metrics, BLEU \cite{papineni2002bleu}, ROUGE \cite{lin2004rouge}, NDCG, HR in sequence prediction problems~\cite{sutskever2014sequence}:
\begin{itemize}[leftmargin=*]
    \item BLEU evaluates the precision of dynamic length recommendations in sequence-to-sequence recommendations~\cite{post2018call}, which is a frequently-used metric in NLP. 
    \item ROUGE refers to the recall of dynamic length recommendations, which is a commonly used metric for sequence-to-sequence recommender systems~\cite{khovrichev2018evaluating}. 
    \item HR@K measures the probability that ground-truth items appear in the top-K recommendations. 
    \item NDCG@K measures the cumulative gain~(CG) scores in the top-K recommendations and considers the influence of position on recommendations~\cite{wang2013theoretical}, where the CG scores represent the similarity between items and ground truths. 
    
    
\end{itemize}

\subsubsection{User Retention.} 
To evaluate the effectiveness of the generated recommendation sequences, we provide two designed metrics, MB-URS and SB-URS, and two common ones, improved user retention (IUR) and no return count (NRC)~\cite{wu2017returning}. 
\begin{itemize}[leftmargin=*]
    \item \textbf{MB-URS}. The \textbf{m}odel-\textbf{b}ased \textbf{u}ser \textbf{r}eturn \textbf{s}core evaluates the effectiveness of recommended item sequences by directly returning a user-retention score. Particularly, we train a supervised model that predicts the reward of particular state-action pairs with the 30\% validation dataset, which is independent of the training dataset. Therefore, the model is optimized by minimizing the MSE loss between predicted rewards and the ground truth. 
    \item \textbf{SB-URS}. The \textbf{s}imilarity-\textbf{b}ased \textbf{u}ser \textbf{r}eturn \textbf{s}core evaluates the effectiveness of recommended item sequences by weighted sum over the ground truth user retention score. Specifically, we divide the samples into 8 classes according to their ground truth reward and calculate BLEU-1 scores between predicted sequences and ground truth for each class as their similarity. Then, the SB-URS are calculated as follows:
    \begin{equation} \small
    \label{equ_sb_urs}
    \text{SB-URS}=\sum_{k=0}^{K}{s_{k} \cdot \left(g_{k}-\frac{K}{2}\right)\cdot N_{k}},
    \end{equation}
    where $s_{k}$ is the similarity of class k. $g_{k}$ is the ground truth reward of class $k$. We want the similarity to be as small as possible for samples with small ground truth reward, and as large as possible for samples with large ground truth reward. $N_{k}$ is the number of samples with a reward of $k$.
    \item \textbf{Improved user retention (IUR)}. 
    It measures the percentage of relative users' average retention improvement compared to their logged average retention in the offline data, which directly reflects users' engagement on the system. And a larger percentage of IUR indicates that recommendation lists make more users engaged in the system.
    \item \textbf{No return count (NRC)}. It is the percentage of users who left the system after a recommendation was made. A smaller no return count indicates that recommendation lists keep more users engaged in the system.
\end{itemize}


\subsection{Baselines}
We compare our model with state-of-the-art methods from different types of recommendation approaches, including:
\begin{itemize}[leftmargin=*]
    \item \textbf{BERT4Rec \cite{sun2019bert4rec}}: It uses a bidirectional Transformer to learn sequential information. In addition, it utilizes a mask language model, to increase the difficulty of the task and enhance the training power of the model.
    \item \textbf{SASRec \cite{kang2018self}}: It employs a left-to-right unidirectional Transformer that captures users' sequential preferences.
    \item \textbf{LIRD \cite{zhao2017deep}}: It combines both policy-based gradient and value-based DQN, using an actor-critic (AC) framework. At each recommendation step, multiple items are recommended, and a simulator automatically learns the optimal recommendation strategy to maximize long-term rewards from users.
    \item \textbf{TopK \cite{chen2019top}}: It uses a policy-based RL approach in offline learning and addresses the distribution mismatch with importance weight. It assumes that the reward of a set of non-repeating items is equal to the sum of rewards of each item and recommends multiple items to the user at each time step. In this paper, we will refer to this model as TopK.
    \item \textbf{DT4Rec-R}: To illustrate the significance of reward in guiding the model, we train a DT model without considering reward, denoted as DT4Rec-R.
\end{itemize}
\begin{table}[t] \center 
    \caption{Overall performance comparison in prediction accuracy. The best performance is marked in bold font.}
	\begin{tabular}{ c  cc  cc c}
		\toprule[1pt]
		\multirow{2}{*}{Dataset} & \multirow{2}{*}{Model} & \multicolumn{4}{c}{Metric}\\
		\cmidrule{3-6}
		& & BLEU$\uparrow$ & ROUGE$\uparrow$ & NDCG$\uparrow$ & HR$\uparrow$ \\
		\midrule
		\multirow{3}{*}{IQiYi} & BERT4Rec & 0.7964 & 0.7693 & 0.7384 & 0.7673\\
		& SASRec & 0.8009 & 0.7906 & 0.7827 & 0.7815\\
		& DT4Rec & \textbf{0.8249*} & \textbf{0.8172*} & \textbf{0.8139*} & \textbf{0.8044*} \\
		\midrule
		\multirow{3}{*}{ML-1m} & BERT4Rec & 0.3817 & 0.3806 & 0.5286 & 0.2769\\
		& SASRec & 0.4052 & 0.3983 & 0.5409 & 0.3123\\
		& DT4Rec & \textbf{0.4331*} & \textbf{0.4185*} & \textbf{0.5679*} & \textbf{0.3342*} \\
		\bottomrule[1pt]
	\end{tabular}
	\\``\textbf{{\Large *}}'' indicates the statistically significant improvements (i.e., two-sided t-test with $p<0.05$) over the best baseline.
	\\ $\uparrow$: the higher the better; $\downarrow$: the lower the better.
	\vspace{-2mm}
	\label{table:result1}
\end{table}

\begin{table}[t] \center 
    \caption{Overall performance comparison in user retention. The best performance is marked in bold font.}
	\begin{tabular}{ c  cc  cc c}
		\toprule[1pt]
		\multirow{2}{*}{Dataset} & \multirow{2}{*}{Model} & \multicolumn{4}{c}{Metric}\\
		\cmidrule{3-6}
		& & MB-URS$\uparrow$ & SB-URS$\uparrow$ & IUR$\uparrow$ & NRC$\downarrow$ \\
		\midrule
		\multirow{4}{*}{IQiYi} & DT4Rec-R & 5.16 & 52131 & 7.5\%  & 2.6\%\\
		& TopK & 5.41 & 61045 & 12.7\% & 2.0\%\\
		& LIRD & 5.63 & 63572 & 17.3\% & 1.9\% \\
		& DT4Rec & \textbf{6.05*} & \textbf{72270*} & \textbf{26.0\%*} & \textbf{1.4\%*} \\
		\midrule
		\multirow{4}{*}{ML-1m} & DT4Rec-R & 5.42 & 13050 & 31.2\% & 9.7\% \\
		& TopK & 5.71 & 13964 & 38.3\% & 6.9\% \\
		& LIRD & 5.86 & 14627 & 41.9\% & 4.8\% \\
		& DT4Rec & \textbf{5.93*} & \textbf{15562*} & \textbf{43.6\%*} & \textbf{3.6\%*} \\
		\bottomrule[1pt]
	\end{tabular}
	\\``\textbf{{\Large *}}'' indicates the statistically significant improvements (i.e., two-sided t-test with $p<0.05$) over the best baseline.
	\\ $\uparrow$: the higher the better; $\downarrow$: the lower the better.
	\vspace{-2mm}
	\label{table:result2}
\end{table}

\subsection{Hyperparameter Setting}
In prompt generator, we set bucket number $B=10$, skip-connection $\alpha=0.1$. In action encoder and decoder, we set maximum sequence length $N=20$, RNN layer number as 1. In Decision Transformer, we set maximum trajectory length T from $\{10, 20, 30, 40, 50\}$, Transformer layer number as 2, heads number as 8, embedding size as 128. We choose AdamW as optimizer and set learning rate as 0.01.
To save the computation cost, we only save the 30 most recently interacted items as the state. For other baselines, the hyper-parameters are set to their optimal values as recommended by their original authors or searched within the same ranges as our model. The results for each model are reported under their optimal hyper-parameter settings. The implementation code is available online\footnote{\url{https://github.com/kesenzhao/DT4Rec.git}}.

\subsection{RQ1: Overall Comparison}
\subsubsection{Prediction Accuracy} 
As shown in Table~\ref{table:result1}, we record all models' best results on the prediction accuracy task. From Table~\ref{table:result1}, we have following observations:
\begin{itemize}[leftmargin=*]
    \item Our model outperforms all baselines on both datasets. This illustrates the effectiveness of our model in optimizing the immediate user feedback. In comparison to the existing methods, our model takes full advantage of RL to model rewards and states of users at each time step. This allows our model to dynamically model the user's interest characteristics to meet the changing needs of the user, so our model can achieve superior performance on the prediction accuracy task.
    \item SASRec performs better than BERT4Rec on both datasets. Our model and SASRec both use a unidirectional Transformer, while BERT4Rec uses a bidirectional Transformer. Since our task is an end-to-end generation task, the unidirectional Transformer is more suitable for our task.
\end{itemize}

\subsubsection{User Retention} 
Table~\ref{table:result2} records all models' best results on maximizing user retention task. We can have following conclusions:

\begin{itemize}[leftmargin=*]
    \item Our model outperforms other baselines on both MB-URS and SB-URS. DT4Rec outperforms the best base LIRD by a large margin on both IQiYi and ML-1m.  
    \item Traditional offline methods, i.e., TopK, LIRD, still suffer from the `Dead Traid' problem and the short-sightedness problem caused by the discount factor since they perform significantly worse than the proposed method. 
    \item In addition, TopK models based on policy gradients are prone to converge to suboptimal solutions, and determining the appropriate learning rate can be challenging. Due to the large action space in SRSs, the LIRD model based on the AC framework is more difficult in the estimation of Q values.
    \item Our model, however, learns rewards and states to action mappings by sequential modeling, eliminating the need for bootstrap and hence avoiding the `Dead Traid' issue. Unlike traditional RL strategies, our model also does not require estimating Q-values. This circumvents the problem of large action space in SRSs. The comparison with DT4Rec-R illustrates the importance of rewards for learning users' long-term preferences.
\end{itemize}

\begin{table}[t] \center
    \caption{Ablation study on IQiYi dataset.}
	\begin{tabular}{@{}c c c c@{}}
		\toprule[1pt]
		\multirow{2}{*}{Architecture} & \multicolumn{3}{c}{Metric} \\ \cmidrule(l){2-4} 
		& SB-URS & MB-URS & IUR \\ \midrule
		DT4Rec & 72270 & 6.05 & 26.0\% \\ 
		w/o contrastive  & 65175 (-9.82\%) & 5.68 (-6.11\%) & 18.3\% (-29.6\%)\\ 
		w/o weight & 69316 (-4.09\%) & 5.79 (-4.30\%) & 20.6\% (-20.8\%)\\ 
		w/o auto-dis & 70953 (-1.82\%) & 5.96 (-1.49\%) & 24.2\% (-6.9\%)\\ 
		\bottomrule[1pt]
	\end{tabular}
	\vspace{-2mm}
	\label{table:ablation}
\end{table}

\subsection{RQ2: Effectiveness of Auto-Discretized Reward Prompt}
By comparing the results of DT4Rec and DT4Rec-R, we have demonstrated the significance of reward as a guide for model training. 
In this subsection, we further show the effectiveness of our improvements for modeling the partial order of reward by ablation study. 
Specifically, we compare DT4Rec with the model using the naive prompt, which is implemented with a single-layer feed-forward neural network, i.e., `w/o auto-dis' in Table~\ref{table:ablation}.
The experimental results show that the auto-discretized reward prompt is effective. Without the auto-discretized reward prompt, the performance on MB-URS and SB-URS drops 1.82\% and 1.49\% respectively, indicating that the auto-discretized reward prompt can make the generated embeddings maintain the partial order relationship between their corresponding reward values.
\begin{figure}[t]
	\centering
	{\subfigure{\includegraphics[width=0.47\linewidth]{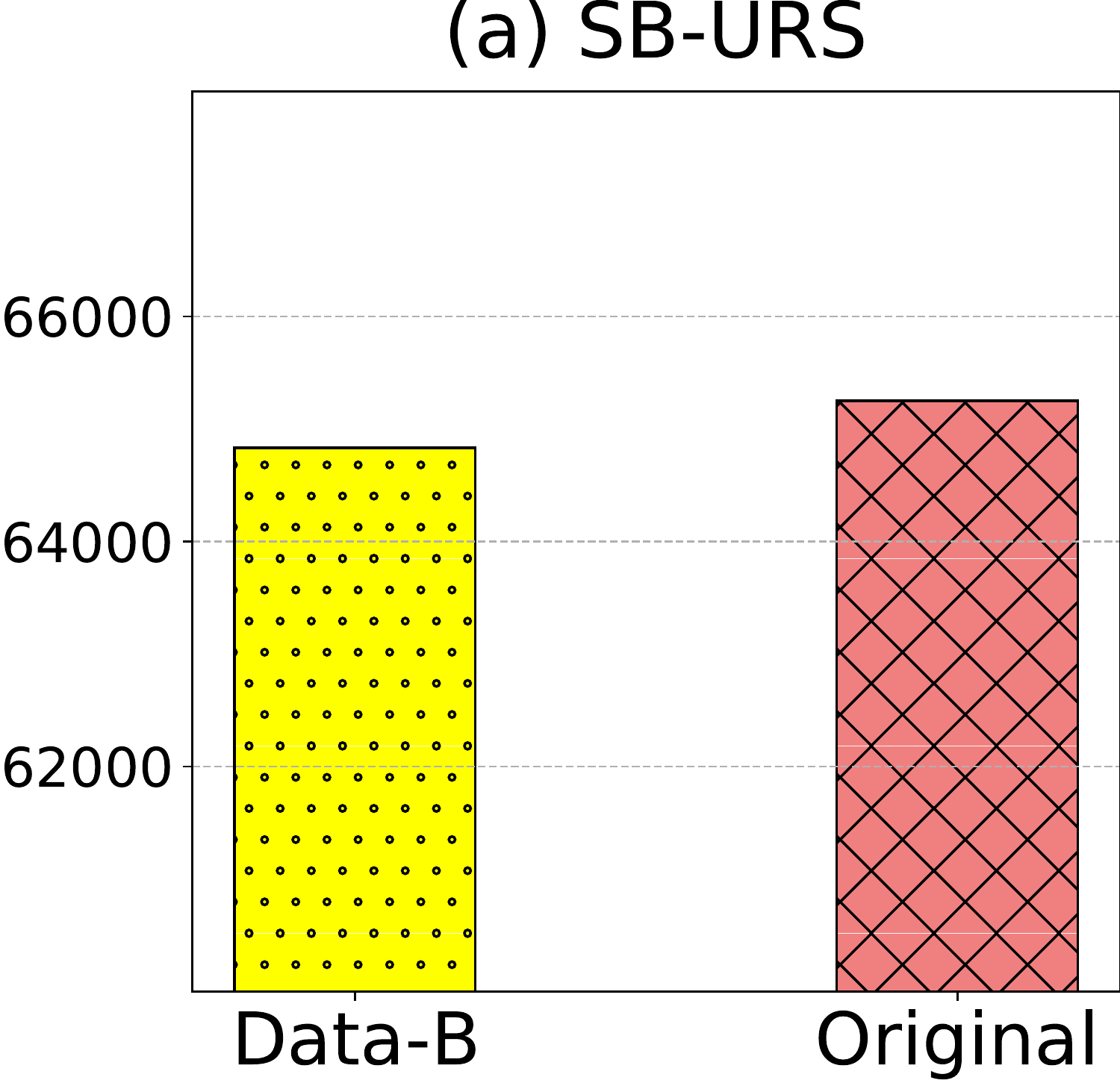}}}
	{\subfigure{\includegraphics[width=0.47\linewidth]{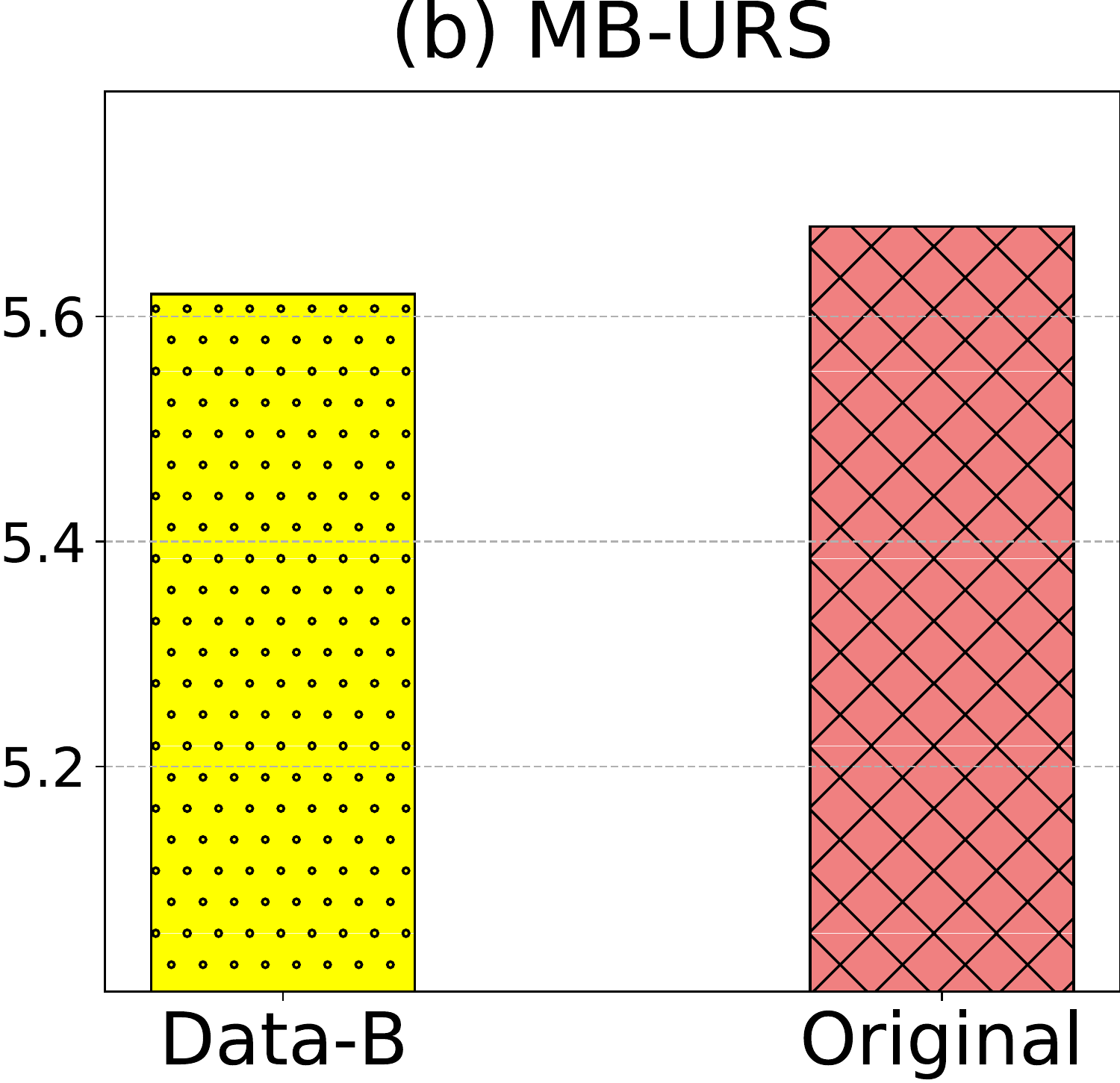}}}
	\vspace{-3mm}
	\caption{Comparison between the model trained on the original IQiYi dataset and the date removing the smaller reward parts, i.e., the Data-B.}\label{fig:data}
	\vspace{-2mm}	
\end{figure}

\begin{figure}[t]
	\centering
	{\subfigure{\includegraphics[width=0.47\linewidth]{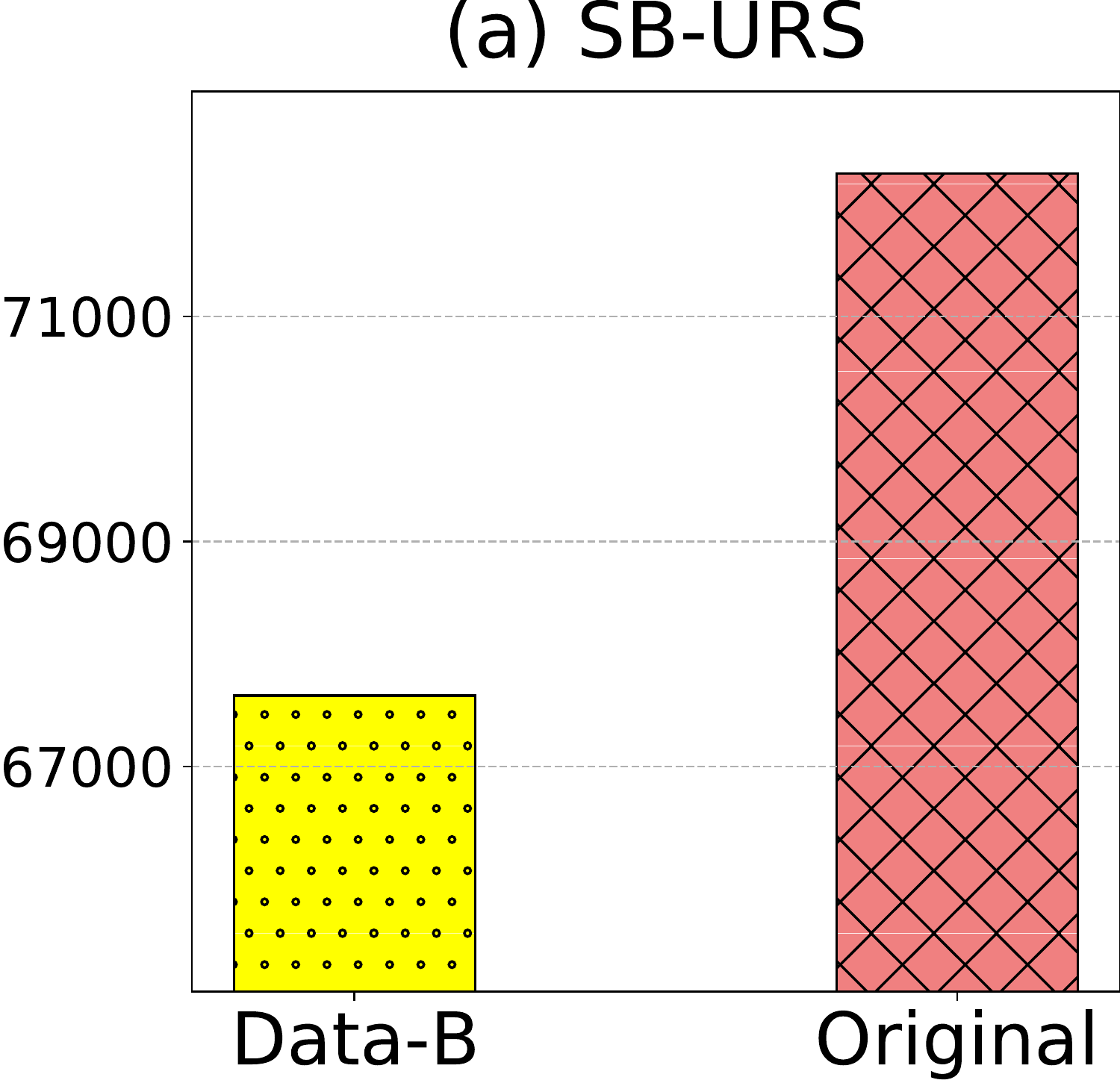}}}
	{\subfigure{\includegraphics[width=0.47\linewidth]{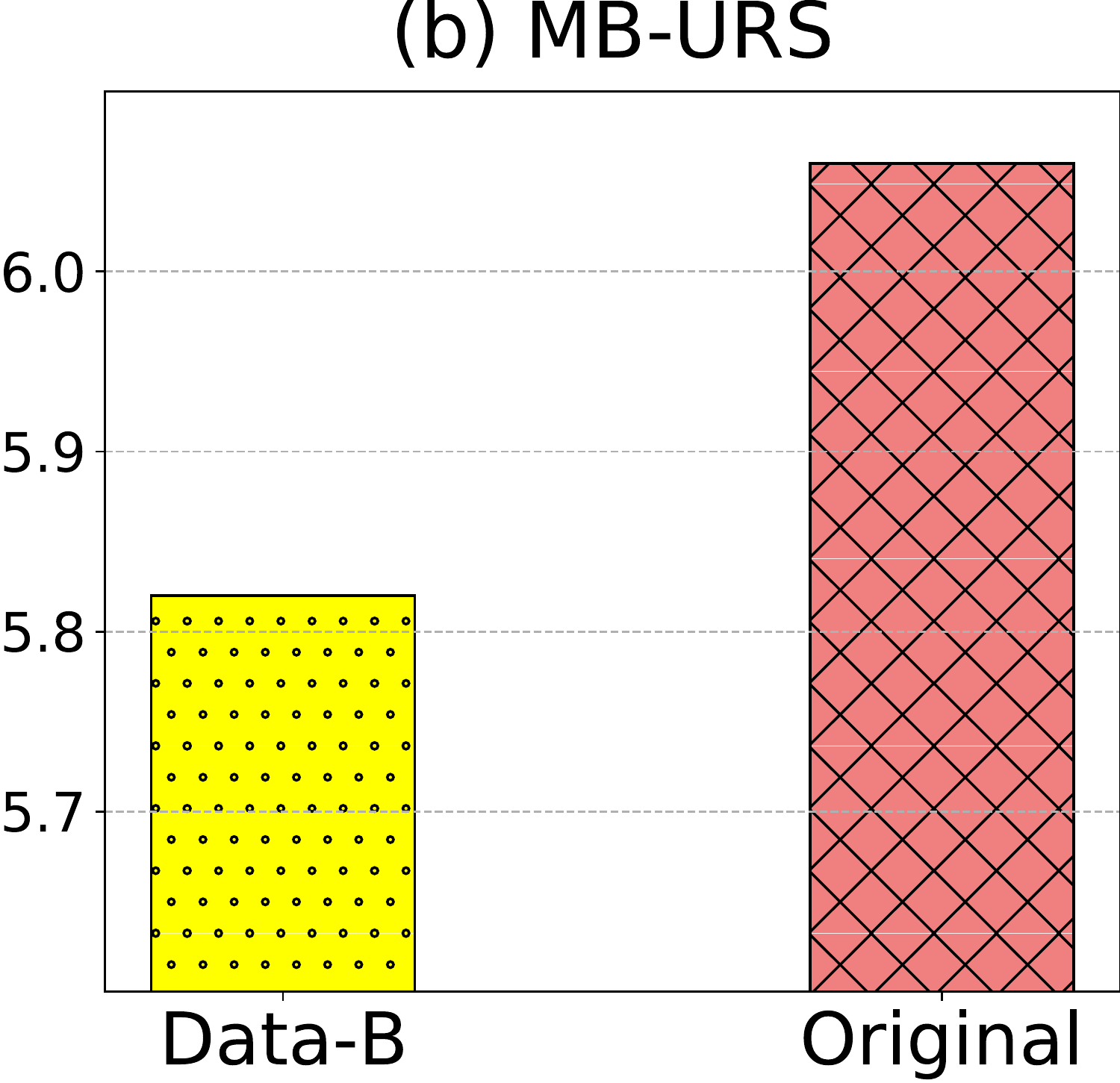}}}
	\vspace{-3mm}
	\caption{Comparison between the DT4Rec trained on the original IQiYi dataset and the date removing the smaller reward, i.e., the Data-B.}\label{fig:comparative}
	\vspace{-2mm}
\end{figure}

\subsection{RQ3: Validity of Contrastive Supervised Policy Learning} \label{sec_ood_analysis}
We further reveal the OOD problems of DT and then powerfully illustrate the effectiveness of our improvements through the following analytical experiments.

\paragraph{Analysis of OOD Problem} 
In the DT model, the model uses the knowledge from the samples with high reward values since only the maximum value of reward is input when generating recommendations. 
Therefore, the knowledge in the samples with small rewards is not incorporated in final recommendation. 
Specifically, we evaluate the model without contrastive learning on the original dataset and data-B, which removes the samples with rewards smaller than 4.
Figure~\ref{fig:data} shows the results on IQiYi dataset.
From Figure~\ref{fig:data}, we can see that there is negligible disparity in the model's performance on the Data-B and the original dataset.
This indicates that the model barely uses the knowledge of samples with a large reward, which verifies our claim on the OOD problem.

\paragraph{Effectiveness of Weighted Contrastive Learning Method~(RQ3)}
In this paragraph, we verify the effectiveness of the weighted contrastive learning from the perspective of improving performance and well-using knowledge of small reward samples. 
For the former, we compare the DT4Rec with the model without the contrast learning loss, denoting it as `w/o contrastive' to demonstrate the superiority of weighted contrastive learning. 
The significant performance drop in Table~\ref{table:ablation} validates the effectiveness of our weighted contrastive learning method. 
For utilizing the knowledge of small reward samples, we compare DT4Rec with the variant that does not use the small reward samples in Figure~\ref{fig:comparative}. 
Particularly, we plot original one and the version without using the small reward samples' performance on SB-URS and MB-URS. 
From the figure, we clearly observe that the SB-URS and MB-URS of our model on original IQiYi dataset are higher than those without using the small reward samples, showing that our model makes full use of the samples with small rewards. 
The reason is that our model will avoid recommending item sequences that are similar to the samples with small rewards, which is consistent with our intuition on weighted contrastive learning method.

\begin{figure}[t]
	\centering
	{\subfigure{\includegraphics[width=0.47\linewidth]{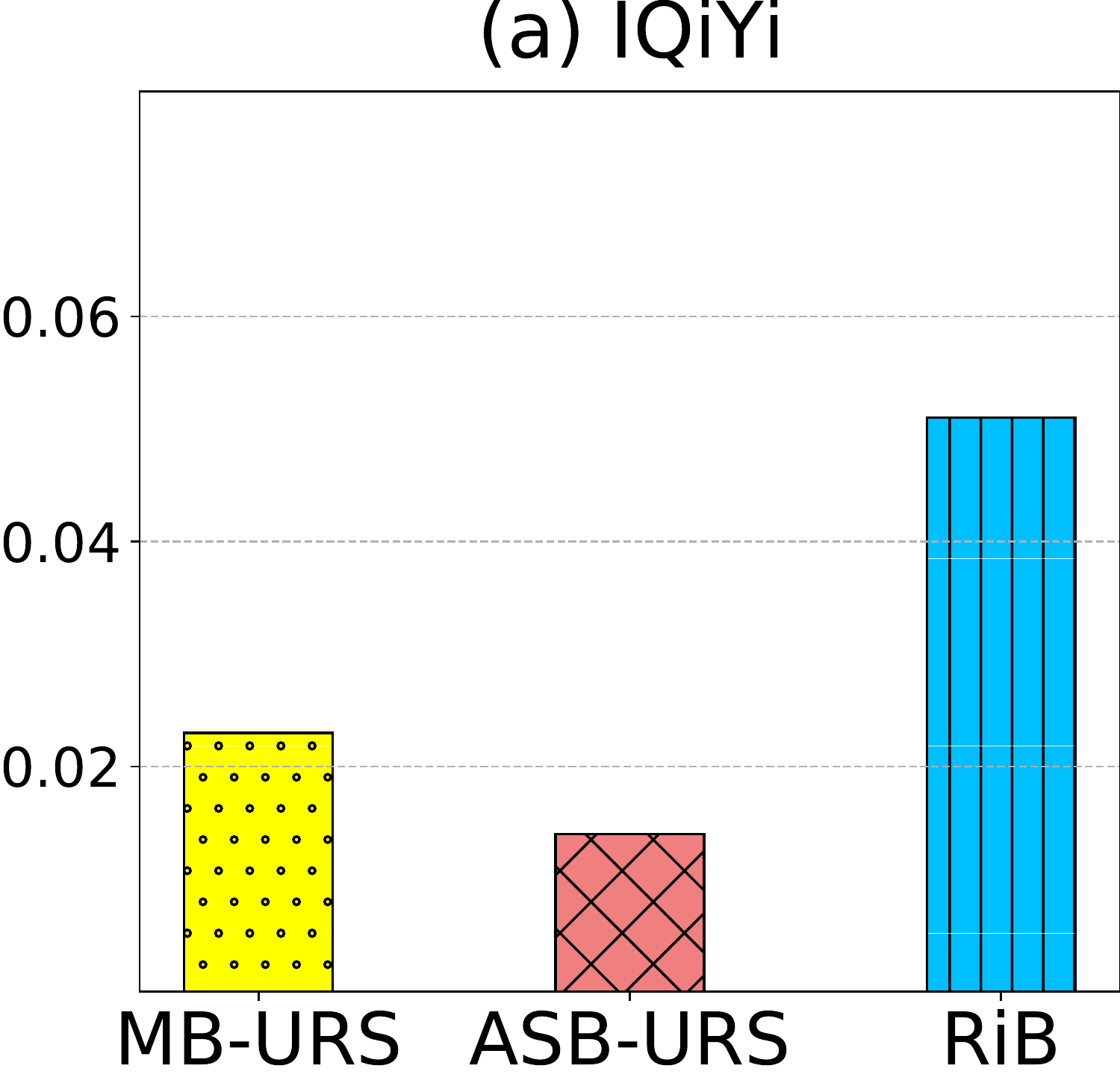}}}
	{\subfigure{\includegraphics[width=0.47\linewidth]{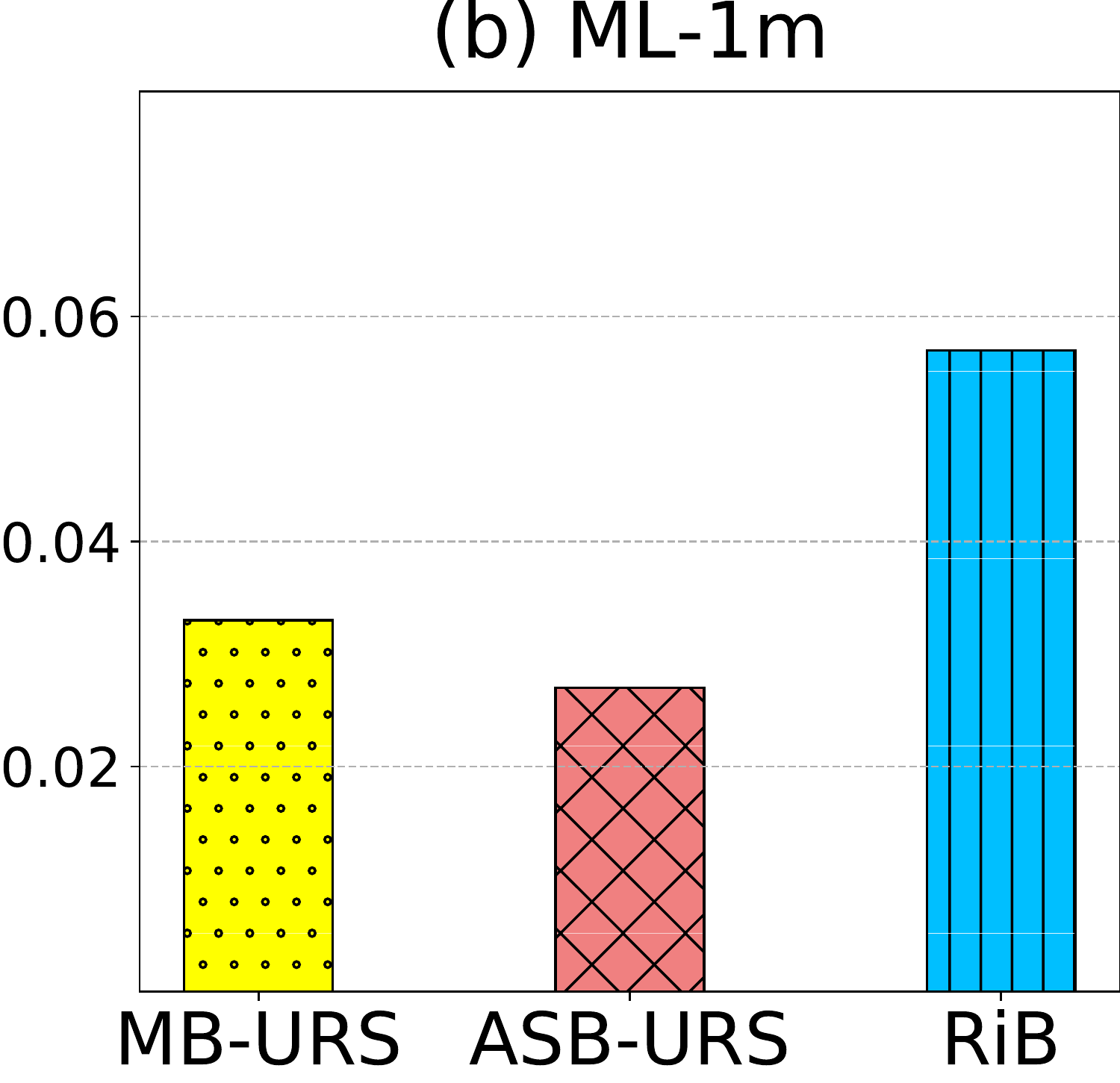}}}
		\vspace{-3mm}
	\caption{Comparison of different evaluation methods' variance on IQiYi and ML-1m datasets.}\label{fig:variance}
\vspace{-3mm}
\end{figure}

\subsection{RQ4: Evaluation Efficiency}
We illustrate the validity of the MB-URS and SB-URS by comparing our methods with the evaluation method proposed by \citeauthor{wu2017returning}, denoting it as `RiB'.
We split the test set into 5 sets and calculate the variance of the performance on these 5 test sets as the indicator of evaluation methods' stability. For fair comparison in the same size, we replace the SB-URS with the ASB-URS, which evaluate the \textbf{a}verage \textbf{s}imilarity-based \textbf{u}ser \textbf{r}eturn \textbf{s}core by dividing the $N_k$ in Equation~\eqref{equ_sb_urs}. 
Figure~\ref{fig:variance} plots the variance of evaluation methods on the IQiYi and ML-1m. The results show that the variance of our evaluation method is small on both datasets, confirming the effectiveness of our proposed methods.

\subsection{RQ5: Model Generalization Analysis}
\begin{figure}[t]
	\centering
	{\subfigure{\includegraphics[width=0.47\linewidth]{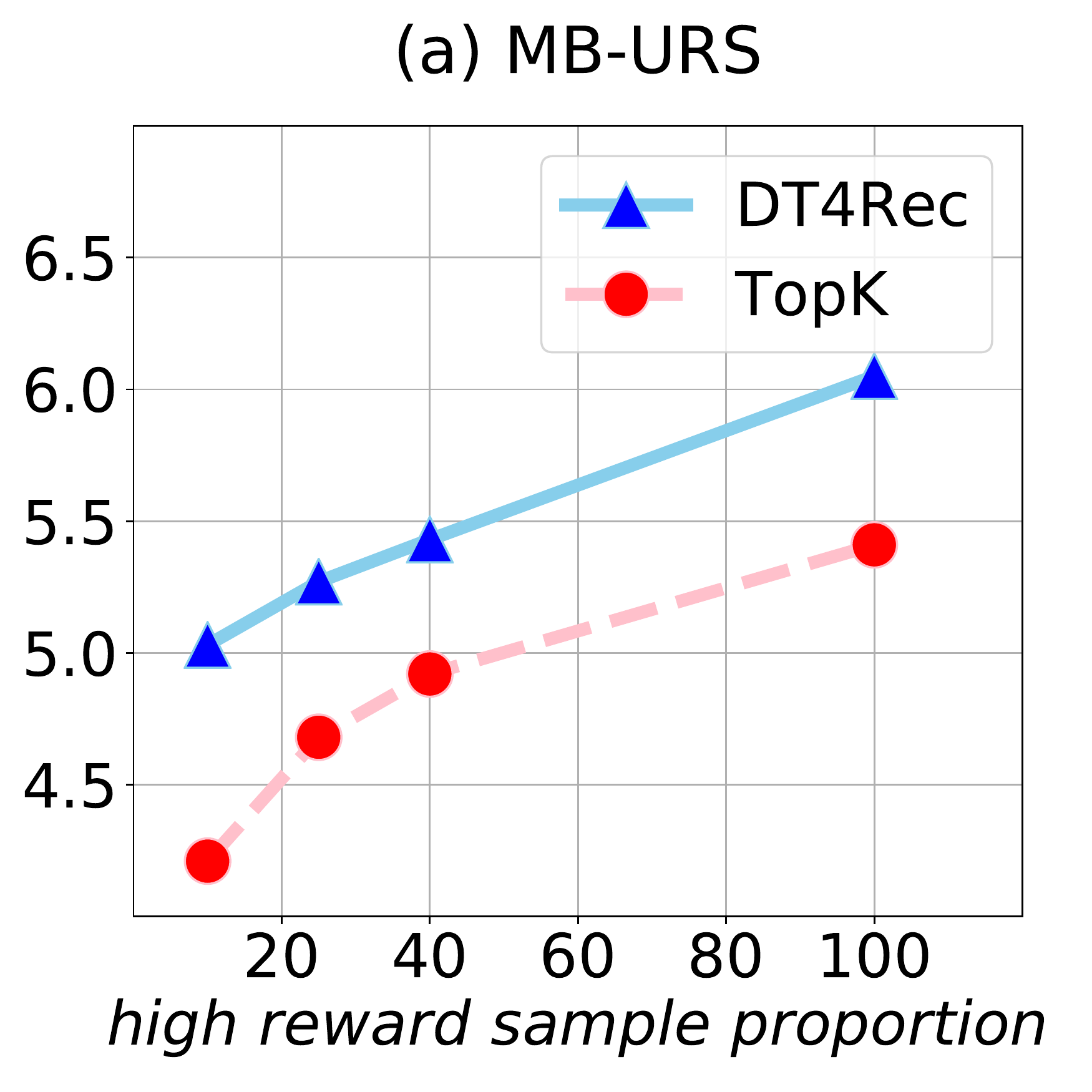}}}
	{\subfigure{\includegraphics[width=0.47\linewidth]{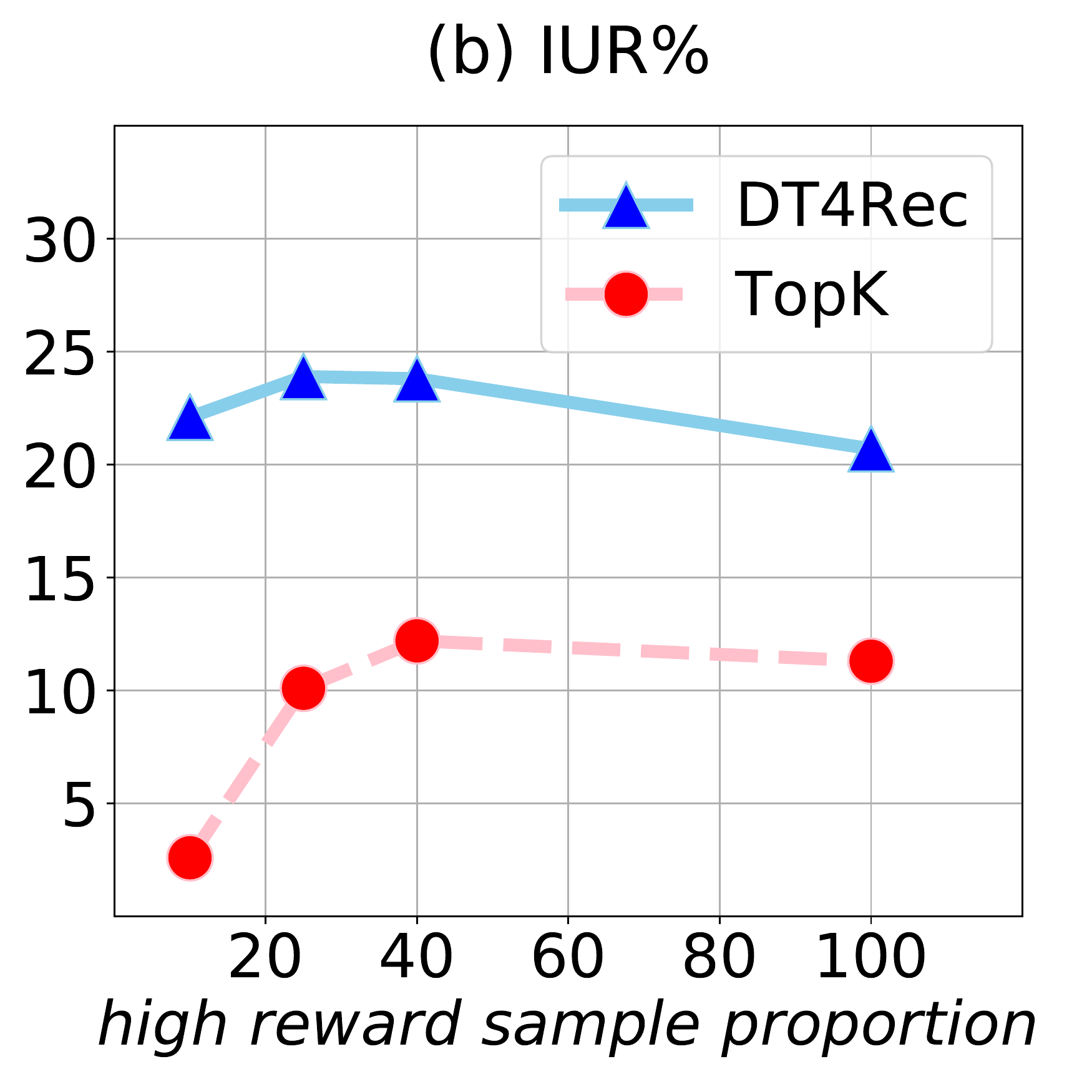}}}
	\vspace{-5mm}
	\caption{Behavior cloning analysis on IQiYi dataset.} \label{fig:BC}
\vspace{-3mm}
\end{figure}

In inference stage, instead of using ground truth reward, only the maximal reward is being utilized for generating the recommendation based on current states. 
Consequently, it leads to the mismatching between training and test since the maximal reward may not exist in the training samples. 
Therefore, it requires the model to have a strong generalization capability, rather than just performing imitation learning on original dataset with a certain return. 
We conduct behavior cloning analysis experiments on IQiYi dataset. Specifically, we treat samples with rewards of 6 or 7 as high reward parts and resplit datasets into high reward parts and not-high reward parts. Furthermore, we control the high reward samples' proportions of the whole dataset to study models' generalization capability. In Figure~\ref{fig:BC}, we plot DT4Rec and TopK's MB-URS and SB-URS on these datasets with different high reward sample proportion, i.e., [10, 25, 40, 100]. From the figure, we observe following facts: (1) on both new datasets, our model performs better than TopK, indicating a strong generalization capability of proposed method. 
(2) DT4Rec's SB-URS is higher than 20\%, even only using 10\% high reward samples. It demonstrates the effectiveness of proposed method in increasing user retention. When the high reward samples is small, the policy evaluation for high reward samples will become inaccurate and hard. However, our model is trained in a supervised way, which avoids the high variance of the policy evaluation and boosts model's performance on test.

%% file: 4Relatedwork.tex
\section{Related Work}
We summarize the earlier literature related to our work, i.e., sequential recommendations and reinforcement learning based recommender systems as follows.
\subsection{Sequential Recommender Systems}
The purpose of sequential recommender systems is to address the order of interactions \cite{li2022mlp4rec, zhang2022hierarchical}.
Early methods are mainly based on Markov Chain and Markov Processes.
For example, 
FPMC learns a transition matrix for each user by combining matrix factorization (MF) and Markov chains (MC) ~\cite{rendle2010factorizing};
Hosseinzadeh et al. capture the variation of user preference by a hierarchical Markov model~\cite{hosseinzadeh2015adapting}. 
In addition to MC-based methods, there are also RNN-based methods.
For example, GRU4Rec~\cite{tan2016improved} firstly proposes to build RS by taking full use of the RNN model to learn sequential information; GRU4RecF proposes the paralleled RNN to leverage item features and improve recommendation efficiency~\cite{hidasi2016parallel}. However, none of these studies are for optimizing the long-term engagement, e.g., user return times~\cite{du2015time}, browsing depth~\cite{zou2019reinforcement}, which is crucial for improving the quality of recommendations.


\subsection{Reinforcement Learning based Recommendation}
Since trial and error search is too costly in recommender systems, here we only present offline reinforcement learning methods. In general, RL-based RSs are mainly divided into two categories: model-based and model-free.

\textbf{Model-based models.} Model-based models
represent the whole environment by learning a fixed model \cite{chen2021survey,zou2020pseudo,zhao2021usersim,zou2021data}. 
However, in recommendation scenario, the model's representation of the environment will be biased due to the dynamic changes in the environment. If the model is constantly updated to adapt to the environmental changes, it will incur a huge expense. In addition, the probability of user's behavior cannot be estimated and the transition probability function becomes difficult to determine. 

\textbf{Model-free models.} In model-free models, the transition probability function is unknown and not required \cite{chen2021survey,zou2020neural}. 
DEERS \cite{zhao2018recommendations} suggests that negative feedback can also have an impact on RSs. RML \cite{montazeralghaem2020reinforcement} improves on past heuristics by directly optimizing metrics for relevant feedback. TPGR \cite{chen2019large} proposes a tree-structured policy-based model to solve the larger discrete action space problem. LIRD \cite{zhao2017deep} utilizes actor-critic framework to make list-wise recommendations. DeepPage \cite{zhao2018deep} further takes into account the 2-D position of the recommended item sequences.
Although model-free models achieve better results than model-based models, many problems remain. First, in RS, due to the large action space, the approximate estimation of value function is very difficult, and the reward function is hard to determine. Second, it is hard to evaluate policy, and the variance of evaluation methods is often unbounded.

%% file: 5Conclusion.tex
\section{Conclusion}
We present a novel reinforcement learning-based sequential recommender system, i.e., DT4Rec, in this work.
Particularly, it avoids the instability problem and unbounded variance headache by casting the RL as an autoregressive problem.
Furthermore, we contribute a series of technologies to the success application of DT4Rec. 
Specifically, the design of an auto-discretized reward prompt effectively models the numerical value of reward and allows guiding the training of models with long-term user engagement.
The proposed contrastive supervised policy learning diminishes the inconsistencies between inference and training of the naive Decision Transformer. 
To evaluate our model, we propose two stable metrics, i.e., MB-URS and SB-URS, which are verified to be more stable than existing ones. 
Extensive experiments conducted on the benchmark datasets have demonstrated the effectiveness of the proposed methods.


\section*{ACKNOWLEDGEMENT}
This research was partially supported by APRC - CityU New Research Initiatives (No.9610565, Start-up Grant for New Faculty of City University of Hong Kong), SIRG - CityU Strategic Interdisciplinary Research Grant (No.7020046, No.7020074), HKIDS Early Career Research Grant (No.9360163), Huawei Innovation Research Program and Ant Group (CCF-Ant Research Fund).